\documentstyle[11pt]{article}
\def\Journal#1#2#3#4{{\em #1} {\bf #2}, #3 (#4)}
\def\Book#1#2#3{{\em #1} ({#2}, {#3})}
\def\Preprint#1#2#3#4{{\em #1}, preprint~{#2}, {#3} (#4)}
\def\JPA{J. Phys. A: Math. Gen.}
\begin{document}
\newcommand{\be}{\begin{equation}}
\newcommand{\ee}{\end{equation}}
\newcommand{\bea}{\begin{eqnarray}}
\newcommand{\eea}{\end{eqnarray}}
\newcommand{\beaa}{\begin{eqnarray*}}
\newcommand{\eeaa}{\end{eqnarray*}}
\newcommand{\qd}{\quad}
\newcommand{\qqd}{\qquad}
\newcommand{\npb}{\nopagebreak[1]}
\newcommand{\nn}{\nonumber}
\newcommand{\dbar}{\bar{\partial}}
\title{\bf Generalized integrable
hierarchies and Combescure symmetry transformations}
\author{L.V. Bogdanov\thanks{Permanent address: IINS,
Landau Institute for Theoretical Physics, Kosygin str. 2,
Moscow 117940, GSP1, Russia; e-mail Leonid@landau.ac.ru}
\hspace{0.1em} and B.G. Konopelchenko\thanks
{Also: Budker Institute of Nuclear Physics, Novosibirsk 90,
Russia}\\
Consortium EINSTEIN\thanks{European Institute for
Nonlinear Studies via Transnationally Extended Interchanges},
\\ Dipartimento di Fisica dell'Universit\`a
and Sezione INFN, \\73100 Lecce, Italy}
\date{}
\maketitle
\begin{abstract}
Unifying hierarchies of integrable equations are discussed.
They are constructed via generalized Hirota identity. It is shown
that the Combescure transformations, known for a long time for the
Darboux system and having a simple geometrical meaning,
are in fact the symmetry transformations of generalized integrable
hierarchies. Generalized equation written in terms of invariants of
Combescure transformations are the usual integrable equations
and their modified partners. The KP-mKP, DS-mDS hierarchies
and Darboux system are considered.
\end{abstract}
\section{Introduction}
The Sato approach (see e.g. \cite{Sato}-\cite{Segal})
and the $\bar{\partial}$-dressing
method (see e.g. \cite{dbar1}-\cite{dbar4})
are two powerful tools
to construct and analyze the hierarchies of integrable equations.
A bridge between these seemingly different approaches
has been established by the observation that the Hirota
bilinear identity can be derived from the $\dbar$-equation
\cite{dbar4}, \cite{Carroll}.
An approach which combines the characteristic
features of both methods, namely, the Hirota bilinear identity from
the Sato approach and the analytic properties of solutions from
the $\dbar$-dressing method, has been discussed in \cite{BK},
\cite{NLS}.

A connection between wave functions with different normalizations
was one of interesting open problems of the $\dbar$-dressing method.
In \cite{BK} it was shown that such a connection is given by the
Combescure transformation.

The Combescure transformation was introduced last century
within the study of the transformation properties of
surfaces (see e.g. \cite{Darboux}, \cite{GEOMA}). It is a transformation
of surface such that the tangent vector at a given point of the surface
remains parallel. The Combescure transformation is essentially
different from the well-known B\"acklund and Darboux transformations.
The Combescure transformation plays an important role in the theory of the
systems of hydrodynamical type \cite{Tsarev}. It is also of great interest
for the theory of (2+1)-dimensional integrable systems \cite{Kon2}.

In the present paper we discuss an approach to integrable
hierarchies based on the generalized Hirota bilinear identity
for the wave function $\psi(\lambda,\mu)$ with simple analytic
properties (Cauchy-Baker-Akhiezer function). We derive generalized
integrable hierarchies in terms of the function $\psi(\lambda,\mu)$.
Such a generalized equation contains usual integrable equations,
their modified partners and corresponding linear problems.
Compact form of the generalized equations
in terms of the $\tau$-function
is derived.

It is shown
that the generalized equations possess the symmetries given by the
Combescure transformations. The invariants of these symmetry
transformations are found. The generalized equations written in
terms of these invariants coincide with the usual equations or
their modified versions. The Darboux transformation and
its connection
with the Combescure transformation for the generalized hierarchies
is also discussed.
The Kadomtsev-Petviashvili (KP) and
modified KP (mKP) equations, the Davey-Stewartson (DS) and
modified Davey-Stewartson (mDS) equations and the matrix
Darboux-Zakha\-rov-Ma\-na\-kov (DZM) system are considered.
\section{Generalized Hirota identity.}
We start with the generalized Hirota bilinear identity
\be
\int_{\partial G} \chi(\nu,\mu;g_1)g_1(\nu)g_2^{-1}(\nu)
\chi(\lambda,\nu;g_2)d\nu=0
\label{HIROTA}
\ee
Here $\chi(\lambda,\mu;g)$ is a matrix function of two complex variables
$\lambda ,\mu\in G$ and a functional of the group element $g$
defining the dynamics (will be specified later), $G$ is some set
of domains of the complex plane, the integration goes over the
boundary of $G$. By definition, the function $\chi(\lambda,\mu)$
possesses the following analytical properties:
$$
{\bar{\partial}}_{\lambda}\chi(\lambda,\mu)=
2\pi {\rm i}\delta(\lambda-\mu),
\quad
-{\bar{\partial}}_{\mu}
\chi(\lambda,\mu)=2\pi {\rm i}\delta(\lambda-\mu),
$$
where $\delta(\lambda-\mu)$ is a $\delta$-function,
or, in other words,
$\chi\rightarrow (\lambda-\mu)^{-1}$
as $\lambda
\rightarrow\mu$ and
$\chi(\lambda,\mu)$ is analytic function of
both variables $\lambda,\mu$ for $\lambda\neq\mu$.

The formula (\ref{HIROTA}) is a basic tool of our construction.

In  another form, more similar to standard Hirota bilinear identity,
the identity (\ref{HIROTA}) can be written as
\be
\int_{\partial G} \psi(\nu,\mu;g_1)
\psi(\lambda,\nu;g_2)d\nu=0 ,
\ee
where
\be
\psi(\lambda,\mu;g)=g^{-1}(\mu)\chi(\lambda,\mu;g)g(\lambda).
\label{substitution}
\ee

We will suggest in what follows that we are able to solve
the equation (\ref{HIROTA}) somehow. Particularly we will bear in mind
that the solutions for equation (\ref{HIROTA}) can be obtained
using the $\bar{\partial}$-dressing method \cite{BK}.
In frame of algebro-geometric technique the function
$\psi(\lambda,\mu)$ corresponds to Cauchy-Baker-Akhiezer kernel
on the Riemann surface (see \cite{Orlov}). The general setting of the problem
of solving (\ref{HIROTA}) requires some modification of
Segal-Wilson Grassmannian approach \cite{Segal}.

Let us consider two
linear spaces $W(g)$ and $\widetilde W(g)$
defined by the function $\chi(\lambda,\mu)$
(satisfying (\ref{HIROTA}))
via equations connected with the identity (\ref{HIROTA})
\bea
\int_{\partial G} f(\nu;g)\chi(\lambda,\nu;g)
d\nu=0,
\label{W}\\
\int_{\partial G}\chi(\nu,\mu;g)h(\nu;g)
d\nu=0 ,
\label{W'}
\eea
here
$f(\lambda)\in W$, $h(\lambda)\in \widetilde W$; $f(\lambda)$,
$h(\lambda)$ are defined in $\bar G$.

It follows from the definition of
linear spaces $W,\;\widetilde W$ that
\bea
f(\lambda)&=&2\pi {\rm i}\int\!\!\!\int_G
\eta(\nu)\chi(\lambda,\nu)d\nu\wedge d\bar{\nu},\quad
\eta(\nu)=\left({\partial\over
\partial\bar\nu}
f(\nu)\right),\nn\\
h(\mu)&=-&2\pi {\rm i}\int\!\!\!
\int_G \chi(\nu,\mu)\widetilde\eta(\nu) d\nu\wedge d\bar{\nu},\quad
\widetilde\eta(\nu)=\left({\partial\over \partial\bar\nu}
h(\nu)\right).
\label{basis}
\eea
These formulae in some sense provide an expansion of the
functions $f\,,h$ in terms of the basic function $\chi(\lambda,\mu)$.
The formulae (\ref{basis}) readily imply that linear spaces
$W,\; \widetilde W$ are transversal to the space of holomorphic functions
in $G$ (transversality property).

From the other point of view, these formulae define a map of
the space of functions (distributions)
on $\bar G$ $\eta,\;\widetilde\eta$
to the spaces
$W$, $\widetilde W$. We will call
$\eta$ ($\widetilde\eta$) a {\em normalization} of the
corresponding function belonging to $W$ ($\widetilde W$).

The dynamics of linear spaces $W,\;\widetilde W$ looks very
simple
\be
W(g)=W_0g^{-1};\quad \widetilde W(g)=g\widetilde W_0,
\label{dynamics}
\ee
here $W_0=W(g=1)$, $\widetilde W_0=\widetilde W(g=1)$
(the formulae (\ref{dynamics})
follow from identity (\ref{HIROTA}) and the formulae (\ref{basis})).

A dependence of the function
$\chi(\lambda,\mu)$ on dynamical variables
is hidden in the function $g(\lambda)$.
We will consider here only the case of continuous variables,
for which
\begin{eqnarray}
g_i&=&\exp(K_i x_i)\label{c}
;\quad
{\partial\over\partial x_i}g=K_i g
\end{eqnarray}
Here $K_i(\lambda)$ are commuting matrix meromorphic
functions.

To introduce a dependence on several variables (may be of different
type), one should consider a product of corresponding functions $g(\lambda)$
(all of them commute).

Let $G$ be a unit disk and $x_{\nu}$ a variable corresponding
to $K_{\nu}(\lambda)={A_{\nu}\over \lambda-\nu}$.
Differentiating the identity (\ref{HIROTA}) over $x_{\nu}$,
one obtains
\bea
{A_{\nu}\over \lambda-\nu}{\partial\over\partial x_{\nu}}
\chi(\lambda,\mu,x_{\nu}) -
{\partial\over\partial x_{\nu}} \chi(\lambda,\mu,x_{\nu})
{A_{\nu}\over \mu -\nu}=
\chi(\nu,\mu)A_{\nu}
\chi(\lambda,\nu),\nn
\eea
or, in terms of the $\psi$ function (\ref{substitution}),
\bea
{\partial\over\partial x_{\nu}} \psi(\lambda,\mu, x_{\nu})=
\psi(\nu,\mu,x_{\nu})A_{\nu}
\psi(\lambda,\nu,x_{\nu}).
\label{N1}
\eea
This formula allows one to construct the basic function
$\chi(\lambda,\mu)$
using
only two functions with `canonical' normalization,
the Baker-Akhiezer function
$\psi(\lambda,\nu,x_{\nu})$ and the dual Baker-Akhiezer
function $\psi(\nu,\mu,x_{\nu})$ corresponding to some
fixed point $\nu$.
\section{The matrix DZM system}
The matrix DZM system is our first example. In this case
the construction and all formulae are very simple and transparent.

To derive the DZM system of equations,
we take a set of three identical unit disks with the center at
$\lambda=0$
${\bf D}_i$, $1\leq i \leq 3$,
as $G$.
The functions $K_i(\lambda)$,  $1\leq i \leq 3$,
are chosen in the form
\bea
K_i(\lambda)={A_i\over \lambda},\quad \lambda\in{\bf D}_i;\nn\\
K_i(\lambda)=0,\quad \lambda\notin {\bf D}_i.\nn
\eea
where $A_i$, $A_j$, $A_k$ are commuting matrices.

It appears that the Hirota bilinear identity in differential
form (\ref{N1}) contains enough information to derive equations
for the rotation coefficients, DZM equations and linear
problem for the DZM equations. Indeed, evaluating the set of
three relations (\ref{N1}) for independent variables
$x_i,\;x_j,\;x_k$ at the set of points
$\lambda,\mu\in \{0_i,\;0_j,\; 0_k\}$
one easily obtains the relations
\bea
\partial_i\psi(\lambda,\mu)&=&\psi(0_i,\mu)A_i\psi(\lambda,0_i),\nn\\
\partial_i\psi(\lambda,0_j)&=&\psi(0_i,0_j)A_i\psi(\lambda,0_i),\nn\\
\partial_i
\psi(0_j,\mu)&=&\psi(0_i,\mu)A_i\psi(0_i,0_j),\nn\\
\partial_i\psi(0_j,0_k)&=&\psi(0_j,0_i)A_i\psi(0_i,0_k),\nn
\eea
where $\partial_i={\partial\over \partial x_i}$.

Let us now take into account that all the equations containing
$\lambda$, $\mu$, can be integrated over the boundary
of $G$ with some matrix weight functions $\rho(\lambda)$,
$\widetilde\rho(\mu)$ (note that they are connected
with the normalization
functions defined by (\ref{basis}))
without changing the structure of equations.
So we will write the equations in terms of
{\em wave functions} independent of spectral parameters
\bea
\partial_i\Phi&=&\widetilde f_i f_i\,,
\label{wave1}\\
\partial_i f_j&=&\beta_{ji}f_i\,,
\label{wave2}\\
\partial_i\widetilde f_j&=&\widetilde f_i \beta{ij}\,,
\label{wave3}\\
\partial_i\beta_{jk}&=&\beta_{ji}\beta_{ik}\,.
\label{rotation}
\eea
Here
\bea
\Phi&=&\int\!\!\!\int \widetilde\rho(\mu)\psi(\lambda,\mu)
\rho(\lambda)d\lambda d\mu,\nn\\
f_i&=&(A_i)^{1\over 2}\int \psi(\lambda,0_i)\rho(\lambda)d\lambda,\nn\\
\widetilde f_i&=&\int \widetilde\rho(\mu)\psi(0_i,\mu)
(A_i)^{1\over 2} d\mu,\nn\\
\beta_{ij}&=&(A_j)^{1\over 2} \psi(0_j,0_i)(A_i)^{1\over 2}.
\label{def}
\eea
The system of equations (\ref{wave1})-(\ref{rotation}) implies
that
\bea
\partial_i\partial_j\widetilde f_k&=&
((\partial_j\widetilde f_i){\widetilde f_i}^{-1})
\partial_i \widetilde f_k
+((\partial_i\widetilde f_j){\widetilde f_j}^{-1}) \partial_j
\widetilde f_k,
\label{ZME1}\\
\partial_i\partial_j \Phi&=&
((\partial_j\widetilde f_i){\widetilde f_i}^{-1})
\partial_i \Phi
+((\partial_i\widetilde f_j){\widetilde f_j}^{-1}) \partial_j
\Phi
\label{ZMEL}
\eea
and
\bea
\partial_i\partial_j f_k&=&
(\partial_i  f_k)({ f_i}^{-1}\partial_j f_i)
+(\partial_j f_k)({ f_j}^{-1}\partial_i f_j),
\label{ZME1d}\\
\partial_i\partial_j \Phi&=&
\partial_i \Phi
{ f_i}^{-1}
(\partial_j f_i)+
\partial_j\Phi
{ f_j}^{-1}
(\partial_i f_j).\label{ZMELd}
\eea
The system (\ref{ZME1}) is just the matrix DZM equation derived in
\cite{dbar1}. The system (\ref{ZME1d}) is its dual partner.
So the solution for the DZM equations
is, in fact, given by the {\em dual wave functions}, i.e. the
wave functions for the linear equations (\ref{wave3}), while
the compatibility conditions for these equations give
the equations for rotation coefficients (\ref{rotation}).
Solutions of the dual DZM system are given by the wave functions
for the linear system (\ref{wave2}).

One always has a freedom to choose
the dual wave function (or, in other words, the freedom to choose
the weight function $\widetilde\rho(\lambda)$), keeping the rotation
coefficients invariant. This freedom is described by the
Combescure symmetry transformation between the solutions
of the DZM system of equations
\be
(\widetilde f'_i)^{-1} \partial_i \widetilde f'_j=
\widetilde f_i^{-1} \partial_i \widetilde f_j
\label{Combescure}
\ee
The equations (\ref{Combescure}) just literally
reflect the invariance of the rotation coefficients.

Similarly for the dual DZM system
\be
(\partial_if'_j)(f'_i)^{-1}  =
(\partial_if_j) f_i^{-1}
\label{CombescureD}
\ee

In fact, the function $\Phi$ is a wave
function for two linear problems (with different potentials),
corresponding to the DZM system
and dual DZM system.
A general Combescure transformation changes solutions
for both the original system and the dual system
(i.e. both functions $\rho(\lambda)$,
$\widetilde\rho(\mu)$).
It is also possible to consider two {\em special}
subgroups of the
Combescure symmetry transformations group.
These two subgroups correspond to the change of
only one weight function, $\rho(\lambda)$ or
$\widetilde\rho(\mu)$; we will call
transformations of this type
right or left Combescure transformations respectively.
The invariants for the right (left) Combescure transformations
are  the solutions
of the dual (or the original) DZM system.
Another form of these invariants is
\bea
(\widetilde f'_i)^{-1} \partial_i \Phi'&=&\widetilde f_i^{-1}
\partial_i \Phi\,,
\label{Combescure+}\\
(\partial_i\Phi')(f'_i)^{-1}&=&(\partial_i\Phi) f_i^{-1}\,.
\label{Combescure-}
\eea

These equations are important
for the connection with the systems of hydrodynamical type
\cite{Tsarev}.
\section{The KP-mKP hierarchy}
The KP-mKP hierarchy is generated by
\be
g({\bf x},\lambda)
=\exp\left(\sum_{i=1}^{\infty}x_i\lambda^{-i}\right)\,,
\label{gKP}
\ee
$G$ in this case is a unit disk.
Let us take
$$
g_1g_2^{-1}=\exp\left(\sum_{i=1}^{\infty}(x_i-x_i')\lambda^{-i}\right)=
\exp\left(-\sum_{i=1}^{\infty}{\epsilon^i\over i \lambda^{i}}\right)=
(1-{\epsilon\over\lambda})\,.
$$
Substituting this function to the Hirota bilinear
identity (\ref{HIROTA}), we get
\bea
&&\left(1-{\epsilon\over\mu}\right)
\chi(\lambda,\mu,{\bf x}')-
\left(1-{\epsilon\over\lambda}\right)
\chi(\lambda,\mu,{\bf x})=
\epsilon \chi(\lambda,0,{\bf x}')\chi(0,\mu,{\bf x})\,,\nn\\
&&x'_i-x_i={1\over i}\epsilon^i\,,
\label{KPbasic0}
\eea
or, in terms of the function $\psi(\lambda,\mu)$
\be
\psi(\lambda,\mu,{\bf x}')-\psi(\lambda,\mu,{\bf x})=
\epsilon \psi(\lambda,0,{\bf x}')\psi(0,\mu,{\bf x});\quad
x'_i-x_i={1\over i}\epsilon^i\,.
\label{KPbasic}
\ee
This equation is a finite form of the whole KP-mKP hierarchy.
Indeed, the expansion of this relation over $\epsilon$ generates
the KP-mKP hierarchies (and dual hierarchies)
and linear problems for them.

Let us take the first three equations given by the
expansion of (\ref{KPbasic}) over~$\epsilon$
\bea
\epsilon:&\;\;&\psi(\lambda,\mu,{\bf x})_x=
\psi(\lambda,0,{\bf x})\psi(0,\mu,{\bf x})\,,
\label{KPbasic1}\\
\epsilon^2:&\;\;&\psi(\lambda,\mu,{\bf x})_y=
\psi(\lambda,0,{\bf x})_x\psi(0,\mu,{\bf x})-
\psi(\lambda,0,{\bf x})\psi(0,\mu,{\bf x})_x\,,
\label{KPbasic2}\\
\epsilon^3:&\;\;&\psi(\lambda,\mu,{\bf x})_t=
{1\over 4}\psi(\lambda,\mu,{\bf x})_{xxx} -
{3\over 4}\psi(\lambda,0,{\bf x})_x\psi(0,\mu,{\bf x})_x+\nn\\
&&{3\over 4}\left ( \psi(\lambda,0,{\bf x})_y\psi(0,\mu,{\bf x})-
\psi(\lambda,0,{\bf x})\psi(0,\mu,{\bf x})_y\right )
\label{KPbasic3} \\
&&x=x_1;\quad y=x_2;\quad t=x_3\,.\nn
\eea
In the order $\epsilon^2$ the equation (\ref{KPbasic})
gives rise equivalently to the equations
\bea
\psi(\lambda,\mu,{\bf x})_y
-\psi(\lambda,\mu,{\bf x})_{xx}&=&-
2\psi(\lambda,0,{\bf x})\psi(0,\mu,{\bf x})_x\,,
\label{KPbasic2+}\\
\psi(\lambda,\mu,{\bf x})_y
+\psi(\lambda,\mu,{\bf x})_{xx}&=&
2\psi(\lambda,0,{\bf x})_x\psi(0,\mu,{\bf x})\,.
\label{KPbasic2-}
\eea
Evaluating the first equation at $\mu=0$, the second
at $\lambda=0$ and integrating them with the weight functions
$\rho(\lambda)$ ($\widetilde\rho(\mu)$), one gets (see (\ref{def}))
\bea
f({\bf x})_y-f({\bf x})_{xx}&=&
u({\bf x})f({\bf x}),\label{LKP}\\
\widetilde f({\bf x})_y+\widetilde f({\bf x})_{xx}&=&-
u({\bf x})\widetilde f({\bf x})\,,
\label{LKPd}
\eea
where $u({\bf x})=-2\psi(0,0)_x$\,.

In a similar manner, one obtains from (\ref{KPbasic1})-
(\ref{KPbasic3}) the equations
\bea
f_t-f_{xxx}={3\over 2}u f_x+
{3\over 4}(u_x+\partial_x^{-1}u_y)f\,,
\label{AKP}\\
\widetilde f_t-\widetilde f_{xxx}={3\over 2}u \widetilde f_x+
{3\over 4}(u_x-\partial_x^{-1}u_y)\widetilde f\,.
\label{AKPd}
\eea
Both the linear system (\ref{LKP}), (\ref{AKP}) for the
wave function $f$ and the linear system (\ref{LKPd}), (\ref{AKPd})
for the
wave function $\widetilde f$ give rise to the same
KP equation
\be
u_t={1\over 4}u_{xxx}+{3\over2}uu_x+{3\over4}\partial_x^{-1}u_{yy}\,.
\ee

To derive linear problems for the mKP and dual mKP
equations, we will integrate equations
(\ref{KPbasic1}), (\ref{KPbasic2+}), (\ref{KPbasic2-})
and (\ref{KPbasic3})
with two weight functions  $\rho(\lambda)$, $\widetilde\rho(\mu)$
(see (\ref{def}))
\bea
\Phi({\bf x})_x&=&
f({\bf x})\widetilde f ({\bf x})\,,\\
\Phi({\bf x})_y-\Phi({\bf x})_{xx}&=&-
2f({\bf x})\widetilde f({\bf x})_x\,,\\
\Phi({\bf x})_y+\Phi({\bf x})_{xx}&=&
2f({\bf x})_x\widetilde f({\bf x})\,,\\
\Phi({\bf x})_t-\Phi({\bf x})_{xxx} &=&-
{3\over2}f({\bf x})_x\widetilde f({\bf x})_x-
{3\over4}(f({\bf x})\widetilde f({\bf x})_y-
f({\bf x})_y\widetilde f({\bf x}))\,.
\eea
Using the first equation to exclude $f$ from the second
(and $\widetilde f$ from the third), we obtain
\bea
\Phi({\bf x})_y-\Phi({\bf x})_{xx}&=&
v({\bf x})\Phi({\bf x})_x\,,
\label{mKPlinear+}\\
\Phi({\bf x})_y+\Phi({\bf x})_{xx}&=&
=-\widetilde v({\bf x})\Phi({\bf x})_x\,,
\label{mKPlinear-}
\eea
where $v=-2{\widetilde f({\bf x})_x\over\widetilde f({\bf x})}$,
$\widetilde v=2{f({\bf x})_x\over f({\bf x})}$.

Similarly, one gets from (\ref{KPbasic3})
\bea
\Phi({\bf x})_t-\Phi({\bf x})_{xxx}&=&
{3\over2}v({\bf x})\Phi({\bf x})_{xx}+
{3\over4}(v_x+v^2+\partial_x^{-1}v_y)\Phi_x\,,
\label{mKPA+}\\
\Phi({\bf x})_t-\Phi({\bf x})_{xxx}&=&
{3\over2}\widetilde v({\bf x})\Phi({\bf x})_{xx}+
{3\over4}(\widetilde v_x+v^2-\partial_x^{-1}\widetilde v_y)\Phi_x\,.
\label{mKPA-}
\eea

The system (\ref{mKPlinear+}), (\ref{mKPA+}) gives rise
to the mKP equation
\be
v_t=v_{xxx}+{3\over4} v^2v_x +3v_x\partial_x^{-1}v_y+
3\partial_x^{-1}v_{yy}\,,
\label{mKP}
\ee
while the system
(\ref{mKPlinear-}), (\ref{mKPA-}) leads to the dual mKP equation,
which is obtained from the (\ref{mKP}) by the substitution
$v\rightarrow\widetilde v$, $t\rightarrow -t$, $y\rightarrow -y$,
$x\rightarrow -x$.

So the function $\Phi$ is simultaneously a wave function
for the mKP and dual mKP L-operators with different potentials,
defined by the dual KP (KP) wave functions.

Using the equation (\ref{KPbasic3}) and relations (\ref{mKPlinear+})
and (\ref{mKPlinear-}), it is possible to obtain an equation for
the function $\Phi({\bf x})$
\bea
\Phi_t-{1\over4}\Phi_{xxx}+{3\over8}
{\Phi_y^2-\Phi_{xx}^2\over\Phi_x}+
{3\over 4}\Phi_x W_y&=&0 ,\quad W_x={\Phi_y\over\Phi_x}\,.
\label{singman}
\eea
This equation first arose in Painleve analysis of the KP
equation as a singularity manifold equation \cite{Weiss}.

The higher analogues of equations (\ref{KPbasic1})-(\ref{KPbasic3})
provide us, with the use of relations
(\ref{mKPlinear+}), (\ref{mKPlinear-}), with the higher analogues
of equation (\ref{singman}). The compact form of the hierarchy of
equations for $\Phi$ can be obtained from the basic finite relation
(\ref{KPbasic}). Integrating both parts of equation (\ref{KPbasic}) with
the weights $\rho(\lambda)$ and $\widetilde\rho(\mu)$, one gets
\be
\Phi({\bf x'})-\Phi({\bf x})-
\epsilon f({\bf x'})\widetilde f({\bf x})=0\,.
\label{singman1}
\ee
Differentiating (\ref{singman1}) with respect to $x_1$, dividing the
result by $ f({\bf x'})\widetilde f({\bf x})$ and using
(\ref{mKPlinear+}), (\ref{mKPlinear-}) , one gets
\be
{\Phi_{x'}({\bf x'})-\Phi_{x}({\bf x})\over \Phi({\bf x'})-\Phi({\bf x})}=
-{\Phi_{y'}({\bf x'})+\Phi_{x'x'}({\bf x'})\over 2\Phi_{x'}({\bf x'})}-
{\Phi_{y}({\bf x})-\Phi_{xx}({\bf x})\over 2\Phi_{x}({\bf x})}\,.
\label{singman3}
\ee

It is also possible to obtain the finite form of KP hierarchy
in terms of the $\tau$-function. Substituting the expression
of the function $\chi(\lambda,\mu)$ through the $\tau$-function
\be
\chi(\lambda,\mu)={\tau\left(g(\nu)\times
\left({\nu-\lambda\over\nu-\mu}\right)\right)
\over \tau(g(\nu))(\lambda-\mu)}
\ee
(see e.g. \cite{NLS}, \cite{Orlov}; $\tau$-function is a functional
of the function $g(\nu)$ or, in other words, a function of ${\bf x}$)
into equation
(\ref{KPbasic0}), one gets
\be
\lambda(\mu-\epsilon)\tau({\bf x}^{(5)})\tau({\bf x}^{(0)})+
\mu(\epsilon-\lambda)
\tau({\bf x}^{(4)})\tau({\bf x}^{(1)})+
\epsilon(\lambda - \mu)
\tau({\bf x}^{(3)})\tau({\bf x}^{(2)})=0\,,
\label{taubasic}
\ee
\bea
&&x_i^{(5)}-x_i^{(4)}=x_i^{(1)}-x_i^{(0)}=
{1\over i}\epsilon^i\,,\nn\\
&&x_i^{(3)}-x_i^{(1)}=x_i^{(4)}-x_i^{(2)}
=-{1\over i}\lambda^i\,,\nn\\
&&x_i^{(5)}-x_i^{(3)}=x_i^{(2)}-x_i^{(0)}={1\over i}\mu^i\,.
\nn
\eea
The expansion of (\ref{taubasic}) in $\epsilon$, $\lambda$,
$\mu$ gives the KP hierarchy in the form of Hirota bilinear
equations.

Equation (\ref{taubasic}) is equivalent to one of the
addition formulae for the $\tau$-function found in
\cite{Sato}.

\section{Combescure transformations for the KP-mKP hierarchy}
Let us consider now the symmetries of the equations
derived above.

Since $\rho(\lambda)$ and $\widetilde\rho(\mu)$ are arbitrary functions,
the equation (\ref{singman}) and the hierarchy (\ref{singman3})
possess the symmetry transformation
$\Phi(\rho(\lambda),\widetilde\rho(\mu))\rightarrow \Phi'=
\Phi(\rho'(\lambda),\widetilde\rho'(\mu))$ .

The Combescure transformation can be
characterized in terms of the
corresponding invariants. The simplest of
these invariants for the mKP equation is just the potential of
the KP equation L-operator expressed through the
wave function
\bea
u&=&{f({\bf x})_y-f({\bf x})_{xx} \over f({\bf x})}\,,
\label{invar1}\\
u&=&{\widetilde f({\bf x})_y-\widetilde f({\bf x})_{xx}
\over \widetilde f({\bf x})}\,,
\label{invar2}
\eea
or, in terms of the solution for the mKP (dual mKP) equation
\bea
v'_y+v'_{xx}-{1\over 2}((v')^2)_x&=&v_y+v_{xx}-{1\over 2}(v^2)_x\,,
\label{invar01}\\
\widetilde v'_y-\widetilde v'_{xx}-{1\over 2}((\widetilde v')^2)_x&=&
\widetilde v_y-\widetilde v_{xx}-{1\over 2}(\widetilde v^2)_x\,.
\label{invar02}
\eea
The solutions of the mKP equations are transformed only by
a subgroup of the Combescure symmetry group corresponding
to the change of the weight function $\widetilde\rho(\mu)$
(left subgroup)
and they are invariant under the action of the subgroup
corresponding to $\rho(\lambda)$ (vice versa for the dual
mKP).

All the hierarchy of the Combescure transformation
invariants is given by the expansion over $\epsilon$
near the point ${\bf x}$ of the
relation (\ref{KPbasic}) rewritten in the form
\bea
{\partial \over \partial \epsilon}
\left ({\widetilde f({\bf x}')-\widetilde f({\bf x})\over
\epsilon \widetilde f({\bf x})}\right ) &=&
-{1\over2}{\partial \over \partial \epsilon}
\partial^{-1}_{x'} u({\bf x}'),\quad
x'_i-x_i={1\over i}\epsilon^i;\\
{\partial \over \partial \epsilon}
\left ({f({\bf x})-f({\bf x'})\over
\epsilon f({\bf x})}\right ) &=&
{1\over2}{\partial \over \partial \epsilon}
\partial^{-1}_{x'} u({\bf x}')\,,\quad
x'_i-x_i=-{1\over i}\epsilon^i.
\eea
The expansion of the left part of these relations gives the
Combescure transformation invariants in terms of the wave functions
$\widetilde f$, $f$. To express them in terms of mKP equation (dual mKP
equation) solution, one should use the formulae
\bea
v&=&-2{{\widetilde f}_x\over \widetilde f}\,,
\quad \widetilde f=\exp(-{1\over2}\partial_x^{-1}v);\\
\widetilde v&=&2{{f}_x\over f},
\quad f=\exp({1\over2}\partial_x^{-1}\widetilde v)\,.
\eea
It is also possible to consider special Combescure transformations
keeping invariant the KP equation (dual KP equation) wave functions
(i.e. solutions for the dual mKP (mKP) equations). The first invariants
of this type are
\bea
{\Phi'_x({\bf x}) \over \widetilde f'({\bf x})}
&=&{\Phi_x({\bf x}) \over \widetilde f({\bf x})}\,,\\
{\Phi'_x({\bf x}) \over f'({\bf x})}
&=&{\Phi_x({\bf x}) \over f({\bf x})}\,.
\eea
All the hierarchy of the invariants of this type is generated
by the expansion of the left part of the following relations over
$\epsilon$
\bea
\left ({\Phi({\bf x}')-\Phi({\bf x})\over
\widetilde f({\bf x})}\right ) &=&
\epsilon f({\bf x}')\,,\quad
x'_i-x_i={1\over i}\epsilon^i;\\
\left ({\Phi({\bf x})-\Phi({\bf x'})\over
f({\bf x})}\right ) &=&
\epsilon \widetilde f({\bf x}')\,,\quad
x'_i-x_i=-{1\over i}\epsilon^i\,.
\eea

Now let us consider the equation (\ref{singman}) and all the
hierarchy given by the relation (\ref{singman3}).
This equation admits the Combescure group of symmetry transformations
$\Phi(\rho(\lambda),\widetilde\rho(\mu))\rightarrow \Phi'=
\Phi(\rho'(\lambda),\widetilde\rho'(\mu))$
consisting of two subgroups (right and left Combescure
transformations). These subgroups have the following invariants
\be
v={\Phi_y-\Phi_{xx}\over\Phi_x}
\label{Comb+}
\ee
and
\be
\widetilde v ={\Phi_y+\Phi_{xx}\over\Phi_x}\,.
\label{Comb-}
\ee
From (\ref{mKPlinear+}), (\ref{mKPlinear-}) it follows that they
just obey the mKP and dual mKP equation respectively.
The invariant for the full Combescure transformation can be
obtained by the substitution of the expression for $v$ via $\Phi$
(\ref{Comb+}) to the formula (\ref{invar01}). It reads
\be
u=\partial_x^{-1}\left({\Phi_y\over\Phi_x}\right)_y-
{\Phi_{xxx}\over\Phi_x} + {\Phi_{xx}^2-\Phi_y^2\over 2 \Phi_x^2}\,.
\label{invarfull}
\ee
From (\ref{LKP}), (\ref{LKPd}), (\ref{invar1}), (\ref{invar2}),
(\ref{invar01}), (\ref{invar02}) it follows that $u$ solves
the KP equation.

So there is an interesting connection between equation (\ref{singman}),
mKP-dual mKP equations and KP equation. Equation (\ref{singman})
is the unifying equation. It possesses a Combescure symmetry
transformations group. After the factorization of equation
(\ref{singman}) with respect to one of the subgroups (right or left),
one gets the mKP or dual mKP equation in terms of the invariants
for the subgroup (\ref{Comb+}), (\ref{Comb-}).
The factorization of equation (\ref{singman})
with respect to the full Combescure transformations group
gives rise to the KP equation in terms of the invariant
of group (\ref{invarfull}).

In other words, the invariant of equation (\ref{singman})
under the full Combescure group is described by the KP equation,
while the invariants under the action of its right and left subgroups
are described by the mKP or dual mKP equations.

Thus the generalized hierarchy (\ref{singman3}) plays a central role
in the theory of the KP and mKP hierarchies.

Using the results of the paper \cite{NLS}, it is possible
to get the formulae for the Darboux type transformation for
the equation (\ref{singman}) in terms of its special solution
$\psi(\lambda,\mu)$. Indeed, a Darboux type transformation
corresponds to
\be
g_d={\nu-b\over\nu-a}\,,\quad a\,,b\in {\bf D}
\label{gDarboux}
\ee
(in fact $a\,,b$ may also belong to regions not connected with
${\bf D}$, this case requires some additional definitions).
The action of $g_d$ (\ref{gDarboux}) on the function
$\chi(\lambda,\mu;g)$
is given by the formula (see \cite{NLS})
\bea
\chi(\lambda,\mu;g\times g_d)=g_d^{-1}(\lambda)g_d(\mu)
{\det\left(\begin{array}{cc}
\chi(\lambda,\mu;g)&\chi(\lambda,a;g)\\
\chi(b,\mu;g)&\chi(b,a;g)
\end{array}\right)\over\chi(b,a;g)}\,.
\eea
In terms of the function $\psi(\lambda,\mu)$ we get
\bea
&&\psi(\lambda,\mu;g\times g_d)=g_d^{-1}(\lambda)g_d(\mu)g(\lambda)
g^{-1}(\mu)\times
\nn\\
&&{\det\left(\begin{array}{cc}
g^{-1}(\lambda)\psi(\lambda,\mu;g)g(\mu)&
g^{-1}(\lambda)\psi(\lambda,a;g)g(a)\\
g^{-1}(b)\psi(b,\mu;g)g(\mu)&
g^{-1}(b)\psi(b,a;g)g(a)
\end{array}\right)\over
g^{-1}(b)\psi(b,a;g)g(a)}
\label{Darboux}
\eea
where $g$ is given by (\ref{gKP})
$$
g({\bf x},\lambda)
=\exp\left(\sum_{i=1}^{\infty}x_i\lambda^{-i}\right)\,.
$$
The formula (\ref{Darboux}) determines a Darboux type
transformation for equation (\ref{singman})
in terms of the function $\psi(\lambda,\mu)$.
We would like to take notice of the fact
that the functions $\psi(\lambda,a;g)$,
$\psi(b,\mu;g)$, $\psi(b,a;g)$ are connected with the function
$\psi(\lambda,\mu;g)$ by the {\em Combescure transformation}
(left, right and their combination).
So the formula (\ref{Darboux}) demonstrates an intriguing connection
between the Darboux and the Combescure transformations for
equation~(\ref{singman}).
\section{Davey-Stewartson - modified Davey-Stewartson hierarchy.
The Ishimori equation}
Now we will consider the 2-component extension of the KP-mKP
hierarchy.
We take a set of two identical unit disks with the center at
$\lambda=0$
${\bf D}_+$, ${\bf D}_-$
as $G$.
The functions $\lambda_+(\lambda)$, $\lambda_-(\lambda)$
are chosen in the form
\bea
\lambda_+(\lambda)=\lambda,\quad \lambda\in{\bf D}_+;&\quad&
\lambda_-(\lambda)=\lambda,\quad \lambda\in{\bf D}_-;
\nn\\
\lambda_+(\lambda)=0,\quad \lambda\in {\bf D}_-;&\quad&
\lambda_-(\lambda)=0,\quad \lambda\in {\bf D}_+\,.
\nn
\eea
The DS-mDS hierarchy is generated by
$$
g({\bf x},\lambda)=
\exp\left(\sum_{i=1}^{\infty}\left (x^+_i\lambda_+^{-i}+
x^-_i\lambda_-^{-i}\right )
\right)\,.
$$
Let us take
$$
g_1g_2^{-1}=
\exp\left(\sum_{i=1}^{\infty}\left(
{\epsilon_+^i\over i \lambda_+^{i}}+{\epsilon_-^i\over i \lambda_-^{i}}
\right)\right)=
(1-{\epsilon_+\over\lambda_+})(1-{\epsilon_-\over\lambda_-})\,.
$$
Substituting this function to the Hirota bilinear
identity (\ref{HIROTA}), we get
\bea
&&\psi(\lambda,\mu,{\bf x}')-\psi(\lambda,\mu,{\bf x})=\nn\\
&&\epsilon_+ \psi(\lambda,0_+,{\bf x}')\psi(0_+,\mu,{\bf x})+
\epsilon_- \psi(\lambda,0_-,{\bf x}')\psi(0_-,\mu,{\bf x}); \nn\\
&&(x^+_i)'-x^+_i={1\over i}\epsilon_+^i,\;
(x^-_i)'-x^-_i={1\over i}\epsilon_-^i\,.
\label{DSbasic}
\eea
The expansion of this relation over $\epsilon_+$,
$\epsilon_-$ generates
the DS-mDS hierarchies (and dual hierarchies)
and linear problems for them.

The DS equation in the usual form is written in terms of
the variables
$\xi=1/2(x+y)=x_1^+$, $\eta=1/2(y-x)=x_1^-$,
$t=-{{\rm i}\over2}(x_2^+ -x_2^-)$.
The DS hierarchy in the form (\ref{DSbasic})
incorporates also the modified Veselov-Novikov hierarchy.

In the standard DS coordinates one gets from (\ref{DSbasic})
\bea
&&\psi(\lambda,\mu,{\bf x})_{\xi}=
\psi(\lambda,0_+,{\bf x})\psi(0_+,\mu,{\bf x})
\label{DSbasic1},\\
&&\psi(\lambda,\mu,{\bf x})_{\eta}=
\psi(\lambda,0_-,{\bf x})\psi(0_-,\mu,{\bf x})
\label{DSbasic2},\\
&&{\rm i}\psi(\lambda,\mu,{\bf x})_t={1\over 2}(
\psi(\lambda,0_+,{\bf x})_{\xi}\psi(0_+,\mu,{\bf x})-
\psi(\lambda,0_+,{\bf x})\psi(0_+,\mu,{\bf x})_{\xi}-\nn\\
&&\psi(\lambda,0_-,{\bf x})_{\eta}\psi(0_-,\mu,{\bf x})+
\psi(\lambda,0_-,{\bf x})\psi(0_-,\mu,{\bf x})_{\eta}).
\label{DSbasic3}
\eea
Just as in the DZM system case, from (\ref{DSbasic1}),
(\ref{DSbasic2}) one obtains the DS and dual DS spatial linear
problems
\bea
\partial_{\eta} f_-=uf_+ &\quad&\partial_{\eta}\widetilde f_-
=v\widetilde f_+
\nn\\
\partial_{\xi} f_+=vf_- &\quad&\partial_{\xi} \widetilde f_+
=u\widetilde f_-
\label{DSL}
\eea
Here $v=\psi(0_-,0_+)$, $u=\psi(0_+,0_-)$.

Similarly to the KP equation case, (\ref{DSbasic3})
gives a time linear problem for the DS equation
\bea
if_{+t}-{1\over 2}f_{+\xi\xi}+{1\over 2}f_{+\eta\eta}&=&
(\partial_{\eta}^{-1}(uv)_{\xi})f_+
-v_{\eta}f_-\,,
\nn\\
if_{-t}-{1\over 2}f_{-\xi\xi}+{1\over 2}f_{+\eta\eta}&=&
-(\partial_{\xi}^{-1}(uv)_{\eta})f_- +u_{\xi}f_+\,.
\label{DSA}
\eea
The compatibility condition for the equations (\ref{DSL}),
(\ref{DSA}) gives the DS equation (in fact it is even easier
to obtain it directly from (\ref{DSbasic1})-(\ref{DSbasic3}))
\bea
iv_{t}-{1\over 2}v_{\xi\xi}-{1\over 2}v_{\eta\eta}&=&
-((\partial_{\xi}^{-1}(uv)_{\eta})+
(\partial_{\eta}^{-1}(uv)_{\xi}))v,\nn\\
iu_{t}+{1\over 2}u_{\xi\xi}+{1\over 2}u_{\eta\eta}&=&
((\partial_{\eta}^{-1}(uv)_{\xi})+(\partial_{\xi}^{-1}(uv)_{\eta}))u.
\label{DS}
\eea

The spatial and time linear problems for the mDS - dual mDS case
read
\bea
&&\Phi_{\eta\xi}=
U_{\xi}\Phi_{\eta} + V_{\eta}\Phi_{\xi}\,,
\label{mDSlinear1}\\
&&{\rm i}\Phi_{t}+{1\over2}\Phi_{\eta\eta}
-{1\over2}\Phi_{\xi\xi} =
V_{\eta}\Phi_{\eta}  -
U_{\xi}\Phi_{\xi}
\label{mDSlinear2}
\eea
and
\bea
&&\Phi_{\xi\eta}=
\widetilde U_{\xi}\Phi_{\eta} +
\widetilde V_{\eta}\Phi_{\xi}\,,
\label{mDSlinear1d}\\
&&{\rm i}\Phi_{t}-{1\over2}\Phi_{\eta\eta}
+{1\over2}\Phi_{\xi\xi} =
-\widetilde V_{\eta}\Phi_{\eta}  +
\widetilde U_{\xi}\Phi_{\xi}\,.
\label{mDSlinear2d}
\eea
Here $V=\log \widetilde f_+$, $U=\log \widetilde f_-$,
$\widetilde V=\log f_+$, $\widetilde U=\log f_-$.
The compatibility
condition for the equations (\ref{mDSlinear1}), (\ref{mDSlinear2})
gives the equations \cite{KonRev}
\bea
\left({\rm i}U_t-{1\over2}U_{\xi\xi}-{1\over2}U_{\eta\eta}-
{1\over2}U_{\xi}^2-{1\over2}U_{\eta}^2+U_{\eta}V_{\eta}\right)_{\eta}
+\left(U_{\eta}V_{\xi}\right)_{\xi}&=&0\,,\nn\\
\left({\rm i}V_t+{1\over2}V_{\xi\xi}+{1\over2}V_{\eta\eta}-
{1\over2}V_{\xi}^2-{1\over2}V_{\eta}^2+U_{\xi}V_{\xi}\right)_{\xi}
+\left(U_{\eta}V_{\xi}\right)_{\eta}&=&0\,.
\label{mDS}
\eea
which can be treated as the modified DS equation.
This system and its connection with
the DS equation has been analyzed in \cite{KonRev}.
The dual
modified DS equation can be
obtained from the (\ref{mDS}) by the substitution
$V\rightarrow\widetilde V$, $U\rightarrow\widetilde U$,
$t\rightarrow -t$, $\xi\rightarrow -\xi$,
$\eta\rightarrow -\eta$.

Solutions for this system are given in terms of dual DS wave functions
(DS wave functions)
Thus there is a Combescure transformations
group acting on the space of solutions. The simplest Combescure
invariants are
\bea
&&{\partial_{\eta}\widetilde f_-\over\widetilde f_+}
=V_{\eta}\exp(V-U)\,,\\
&&{\partial_{\xi}\widetilde f_+\over\widetilde f_-}
=U_{\xi}\exp(U-V)\,.
\eea
The hierarchy of the Combescure transformation invariants
is generated by the relations
\bea
&&\left ({\widetilde f_+({\bf x}')-\widetilde f_+({\bf x})\over
\widetilde f_-({\bf x})}\right ) =
\epsilon_+ u({\bf x'})\,,\\
&&(x^+_i)'-x^+_i={1\over i}\epsilon_+^i;\quad
x^-_i-(x^-_i)'=0;\nn\\
&&\left ({\widetilde f_-({\bf x}')-\widetilde f_-({\bf x})\over
\widetilde f_+({\bf x})}\right ) =
\epsilon_-v({\bf x'})\,,\\
&&(x^-_i)'-x^-_i={1\over i}\epsilon_-^i;\quad
(x^+_i)'-x^+_i=0\,.\nn
\eea
If one takes a pair of wave functions $\widetilde f_+,\widetilde f_-$
and $\widetilde f'_+,\widetilde f'_-$, the matrix
$$
\Psi=\left(
\begin{array}{cc}
\widetilde f_+&\widetilde f'_+\\
\widetilde f_-&\widetilde f'_-
\end{array}\right)
$$
is connected with the solution of the Ishimori equation by the formula
(see e.g. \cite{KonRev})
$$
S_1\sigma_1+S_2\sigma_2+S_3\sigma _3=-\Psi^{-1}\sigma_3\Psi
$$
(for real ${\bf S}$ some reduction conditions should be
satisfied). In principle it could be possible to express
Combescure invariants for mDS equation in terms of
solution for the Ishimori equation and thus obtain Combescure
invariants for the Ishimori equation, but it is unclear in this case
whether the Combescure transformation survives under
reduction conditions.
\subsection*{Acknowledgments}
The first author (LB) is grateful to the
Dipartimento di Fisica dell'Universit\`a
and Sezione INFN, Lecce, for hospitality and support;
(LB) also acknowledges partial support from the
Russian Foundation for Basic Research under grant
No 96-01-00841.

\end{document}